\def\ontop#1#2{\begin{array}{c}
               #1 \\ #2
               \end{array}}
\documentclass[]{cupconf}
\usepackage{graphicx}

\title[Massive star evolution at high metallicity]
      {Massive star evolution at high metallicity}
      
\author[G.~Meynet et al.]{\\ Georges Meynet$^1$, Nami Mowlavi$^2$ and Andr\'e Maeder$^1$} 
\affiliation{$^1$ Observatoire de Gen\`eve, Universit\'e de Gen\`eve, CH--1290 Sauverny, Switzerland \\ $^2$ ISDC, Observatoire de Gen\`eve, Universit\'e de Gen\`eve, Chemin d'Ecogia 16, CH-1290 Versoix}   

\begin{document}
\maketitle

\begin{abstract}
After a review of the many effects of metallicity on the evolution of rotating and non-rotating stars, we discuss the consequences of a high metallicity on massive star populations and on stellar nucleosynthesis. 
The most striking effect of high metallicity is to enhance the amount of mass lost by stellar winds.
Typically at a metallicity $Z=0.001$ only 9\% of the total mass returned by non-rotating massive stars is ejected by winds
(91\% by supernovae explosion), while at solar metallicity this fraction may amount to more than 40\%.
High metallicity favors the formation of Wolf-Rayet stars and of type Ib supernovae. It however disfavors the occurrence
of type Ic supernovae. We estimate empirical yields of carbon based on the observed population of WC stars in the solar neighborhood, and obtain that WC stars eject between 0.2 and 0.4\% of the mass initially locked into stars under the form of new synthesized carbon. Models give values well in agreement with these empirical yields.
Chemical evolution models indicate that such carbon yields may have important impacts on the abundance
of carbon at high metallicity.
\end{abstract}

\firstsection
\section{General effects of metallicity on the evolution of stars}

The metallicity affects the evolution of stars mainly through its
impact on the radiative opacities, the equation of state, the nuclear
reaction rates, and the stellar mass loss rates (see also the discussion by Maeder 2002). Metallicity also affects the way
the various instabilities induced by rotation occur in stars. 

Recent calculations ordered by increasing metallicities are briefly presented in Table~\ref{t1},
the first column gives the reference, the second to fourth columns indicate respectively
the range of initial masses, the values of the initial mass fraction of helium and the value
of $Z$ considered in the different grids of stellar models. 

In this paper we first review the various effects of metallicity on the evolution of stars, then
we discuss the consequences of high metallicity on the massive star populations and their associated nucleosynthesis.

\subsection{The $\Delta Y/\Delta Z$ ratio}

let us recall how a change of $Z$, the mass fraction of heavy elements, affects
the initial mass fraction of hydrogen, $X$,  and of helium, $Y$.
In the course of galactic
evolution, the interstellar matter is progressively enriched in heavy elements and in helium.
The increase of the abundance in helium with respect to $Z$ is conveniently described by a
$\Delta Y/\Delta Z$ law such that $Y=Y_0+\Delta
Y/\Delta Z\times Z$ (where $Y_0$ is the primordial He abundance). At low metallicities
($Z\leq 0.02$),
$Y$ remains practically constant, and the
stellar properties {\it as a function of metallicity} are thus expected to be
determined mainly by $Z$. At high metallicities ($Z\geq 0.02$), on
the other hand, $Y$ increases (or, alternatively, $X$ decreases) significantly
with $Z$. In those conditions, both $Z$ {\it and} $Y$ (and $X=1-Y-Z$)
determine the stellar properties as a function of metallicity.

\begin{table}
\caption{Grids of high metallicity massive star models.}
\begin{center}
\begin{tabular}{lccr}
 & & & \\
Reference & Initial Masses & $Y$ & $Z$ \\ 
 & & & \\
Claret 1997 & 1 - 40 & 0.42 - 0.32 - 0.22 & 0.03 \\
Schaerer et al. 1993& 0.8 - 120 & 0.340 & 0.04 \\ 
Meynet et al. 1994 & 12 - 120 & 0.320 & 0.04 \\
Meynet \& Maeder 2005$^1$ & 20 - 120 & 0.310 & 0.04 \\
Eldridge \& Vink 2006$^2$ & 25 - 120 & 0.340 & 0.04 \\
Fagotto et al. 1994a & 0.6 - 120 & 0.352 & 0.05 \\
Fagotto et al. 1994b & 0.6 - 9   & 0.475 & 0.10 \\
Mowlavi et al. 1998a & 0.8 - 60 & 0.480 & 0.10 \\
 & & & \\
\multispan{4}{$^1$ Models including the effects of rotation\hfill.}\\
\multispan{4}{$^2$ Models with Z-dependent mass loss rates during the WR phase.}\\

\end{tabular}
\end{center}
\label{t1}
\end{table}

\subsection{The $\kappa$-effect}
\label{Sect:kappa-effect}

\begin{figure}
\includegraphics[width=2.6in,height=2.6in]{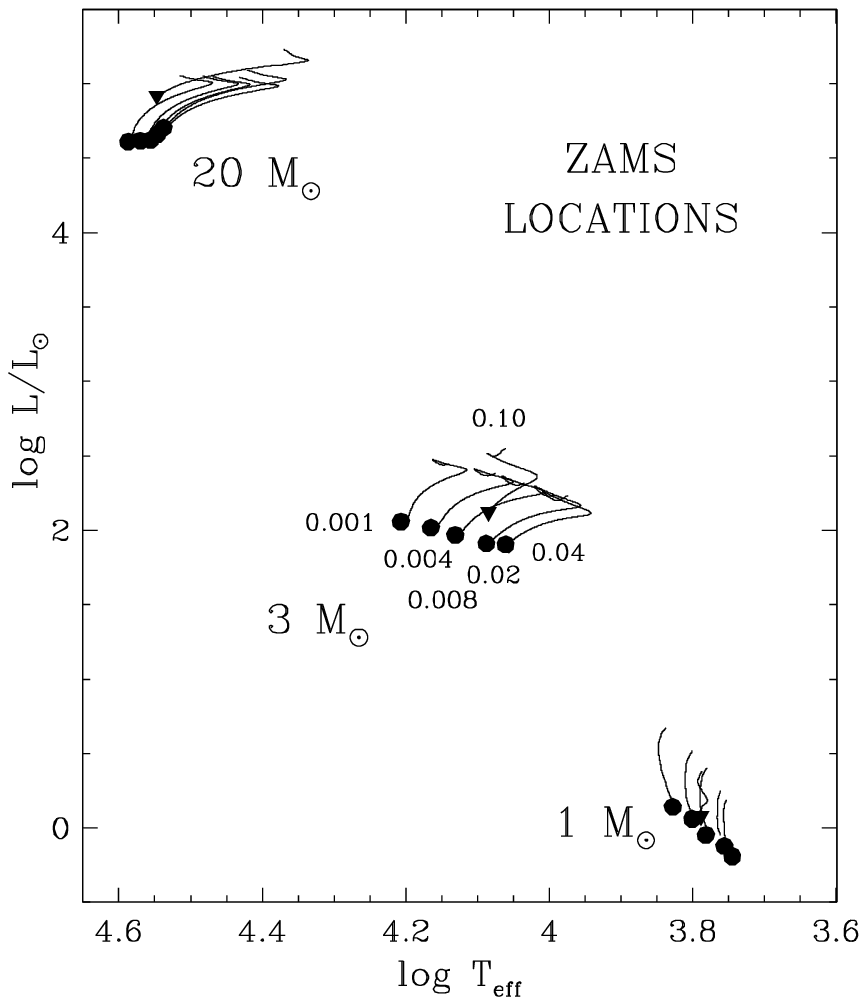}
\hfill
\includegraphics[width=2.77in,height=2.77in]{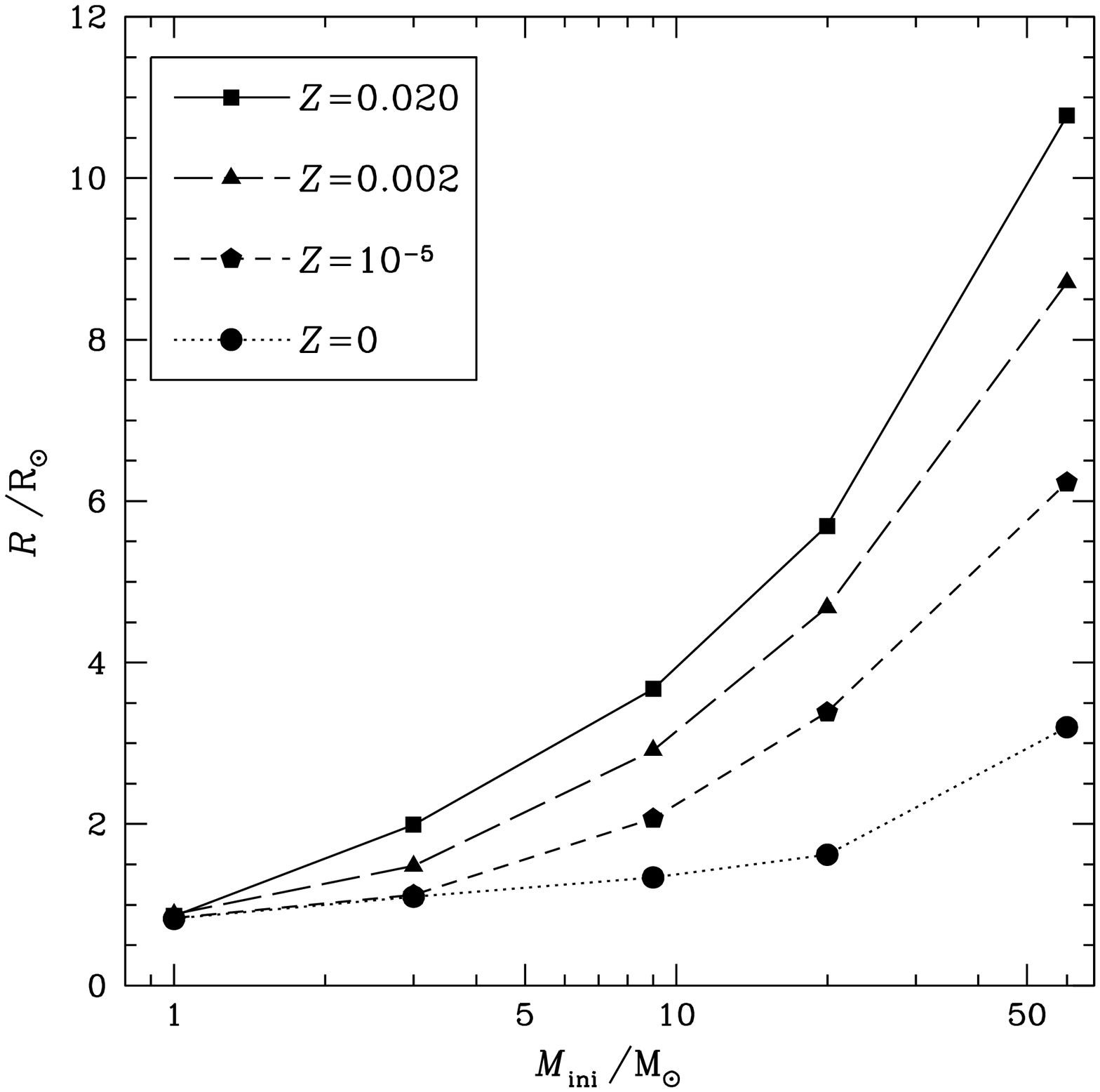}
\caption{{\it Left panels}:ZAMS locations in the HR diagram of 1, 3 and 20 M$_\odot$ star
           models at different metallicities. Models at $Z=0.1$ are identified by triangles (Figure taken from Mowlavi et al. 1998b).{\it Right panel}: Variations of the radius as a function of initial mass, for various metallicities on the ZAMS.}\label{fig1}
\end{figure}

In general, the opacities increase with increasing metallicity.
This is the case for the bound-free (see Eq.~ 16.107 in Cox \& Giuli 1969)
and free-free (see Eq.~16.95 in Cox \& Giuli 1969) transitions.
Using the well known mass-luminosity relation $L \propto \sim {\mu^4 M^3 \over \kappa}$, where $\mu$ is the mean molecular weight, $M$ the mass and $\kappa$ the opacity, one immediately deduces that the increase
of the opacity produces a decrease of the luminosity. This can be seen in the left part of Fig.~\ref{fig1} for the 3 and 1 M$_\odot$ stellar models and for metallicities inferior
or equal to 0.04. At still higher metallicities the effect of $\mu$ becomes more important
than the opacity effect and the luminosity increases (see below).

In contrast, for massive stars free electron scattering is the main opacity source. 
This opacity depends only on $X$
[$\kappa_e\simeq 0.20(1+X)$], which is about constant at
$Z\leq 0.01$ and decreases with $Z$ at $Z \geq 0.01$.
Thus at high $Z$ the luminosity of the 20 M$_\odot$ model shown in the left part of Fig.~\ref{fig1} first remains nearly constant and then increases with the metallicity.

In the right part of Fig.~\ref{fig1} the variation of the radii on the ZAMS for stars of various initial masses and metallicities are shown. We see that the radii increase with the metallicity. From the relation $L=4\pi R^2\sigma T^4_{\rm eff}$ and from the variation of $L$ with $Z$, one can deduce
that when the metallicity increases, the effective temperature decreases. For metallicities superior to about 0.04, the effective temperature and the luminosity both 
increase due to the $\mu$-effect (see below).


\subsection{The $\mu$-effect}
\label{Sect:mu-effect}

When the metallicity increases, $\mu$ increases. This tends to make the star more luminous
according to the mass-luminosity relation, provided the opacity does not increase too much. At very high metallicity (Z $> \sim 0.04$), this is what happens for the 1 and 3 M$_\odot$ models.
For the 20 M$_\odot$ stellar mass model, the $\mu$-effect is also present as can be seen from Fig.~\ref{fig1} (left part).The luminosity already begins to increase as a
function of $Z$ at $Z\simeq 0.02$ since the $\kappa$-effect on $L$ is negligible.

\subsection{The $\varepsilon_{nuc}$-effect}
\label{Sect:epsilon-effect}

 The nuclear energy production $\varepsilon_{nuc}$ sustains the stellar
luminosity. If $\varepsilon_{nuc}$ is arbitrarily increased, the central regions
of the star expand, leading to a decrease in the temperatures and densities, and
to an increase in the stellar radius. The temperature gradient therefore decreases
leading to a decrease of both the luminosity and of the effective temperature.

The way $\varepsilon_{nuc}$ depends on  metallicity is closely related to the
mode of nuclear burning.
When the CNO cycles are the main mode of H burning, $\varepsilon_{nuc}$ increases
with $Z$.
In contrast, when the main mode of burning is the pp chain, $\varepsilon_{nuc}$
is related to $X$. It is thus about independent of $Z$ at $Z\leq 0.02$, and
decreases with increasing $Z$ at $Z\geq 0.02$.
These properties determine the behavior of
$T_c$ and $\rho_c$ in MS stars as a function of $Z$.

Usually, the CNO cycles are the main mode of core H burning
for stars more massive than about 1.15-1.30 M$_\odot$ (depending on metallicity),
while the pp chain operates at lower masses. At $Z=0.1$, however, all stars
are found to burn their hydrogen through the CNO cycles (at least down to
$M=0.8$ M$_\odot$ see Mowlavi et al. 1998a).

\begin{figure}
\includegraphics[width=2.6in,height=2.6in]{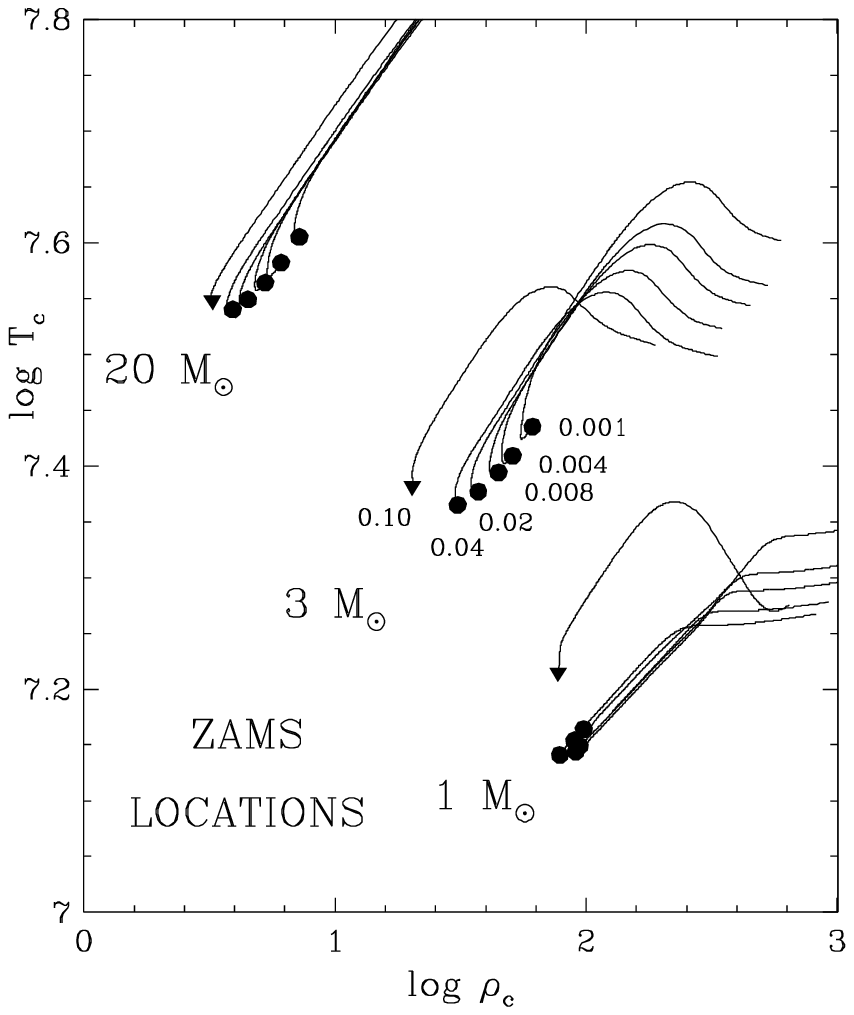}
\hfill
\includegraphics[width=2.6in,height=2.6in]{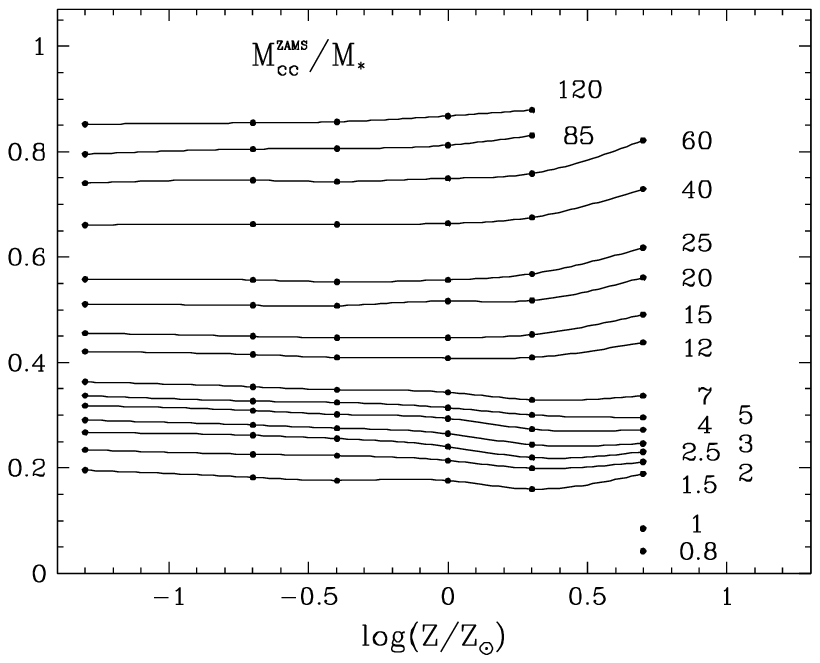}
\caption{{\it Left panels}: Same as the left part of Fig.~\ref{fig1} but in the ($\log \rho_c, \log T_c$) diagram.{\it Right panel}: Masses of convective cores relative to the stellar masses
           as labeled (in solar masses) next to the curves, as a function
           of metallicity. Figures taken from Mowlavi et al. (1998b).}\label{fig2}
\end{figure}

 When the CNO cycle is the main mode of burning, $\rho_c$ and $T_c$ decrease with
increasing $Z$ (see above).
This is clearly visible
in the left part of Fig.~\ref{fig2} for the 3 and 20 M$_\odot$ stars at $Z\leq 0.04$.
At $Z=0.1$, however, the higher surface luminosities of the models as compared to
those at $Z=0.04$ result in concomitant higher central temperatures.

 When the pp chain is the main mode of burning, on the other hand,
the location in the $(\log \rho_c,\log T_c)$ is not very sensitive to $Z$.
This is well verified for the 1 M$_\odot$ models at $Z<0.1$.

The combined effects of $\mu$, $\kappa$ and $\varepsilon_{nuc}$ on the
stellar surface properties can be estimated more quantitatively with a
semi-analytical approach using homology relations. This is developed in the appendix of 
Mowlavi et al. (1998b). We just recall below that a variation $\Delta \mu$ of the mean molecular weight of a star of mass $M$ affects
its position in the HR diagram by

\begin{eqnarray}
  \label{Eq:DeltaL0}
  \Delta \log L =   \left\{\ontop{7.3}{7.8}\right\} \Delta \log \mu
                  - \left\{\ontop{1.02}{1.08}\right\} \Delta \log \kappa_0
                  - \left\{\ontop{0.02}{0.08}\right\} \Delta \log \epsilon_0
\end{eqnarray}
\begin{eqnarray}
  \label{Eq:DeltaTeff0}
  \Delta \log T_{eff} =   \left\{\ontop{1.6}{2.2}\right\} \Delta \log\mu
                     - \left\{\ontop{0.28}{0.35}\right\} \Delta \log\kappa_0   
                     - \left\{\ontop{0.03}{0.10}\right\} \Delta \log\epsilon_0
\end{eqnarray}
where $\epsilon_0$ the
temperature
and density independent coefficient in the relation for the nuclear energy
production $\epsilon_{nuc}=\epsilon_0\;\rho^\lambda\;T^\nu$ (see Cox and Giuli
1969, p~692), and $\kappa_0$ the
opacity coefficient in the Kramers law $\kappa=\kappa_0\;\rho\;T^{-3.5}$.
The numbers on the first line in Eqs.~\ref{Eq:DeltaL0} and
\ref{Eq:DeltaTeff0}
apply when the nuclear energy production results from the $pp$ chain, while the
second
line applies when the $CNO$ cycles provide the main source of nuclear energy.
Variation of $\mu$ is related to variation of $Z$ by
$\Delta \log \mu= \frac{4.5}{\ln 10} \mu \Delta Z$.

The mass of the convective core is mainly determined by the nuclear energy production,
and by $L$.  At $Z\leq 0.04$, it
slightly decreases with increasing $Z$ in low- and intermediate-mass stars, while it
remains approximately constant in massive stars (see right part of Fig.~\ref{fig2}).
At $Z\geq 0.04$, on the other hand, it increases with $Z$ at all stellar
masses\footnote{It is interesting to note that we could have expected smaller
convective cores in metal-rich massive stars where the main source of opacity is
electron scattering (since this opacity is positively correlated with the H content). But
the effect of higher luminosity in those stars overcomes that $\kappa$-effect, and
the core mass increases with increasing $Z$.}.
We recall furthermore that at $Z=0.1$ all stars with $M$ as low as 0.8 M$_\odot$
possess convective cores.

\begin{figure}
\includegraphics[width=2.6in,height=2.6in]{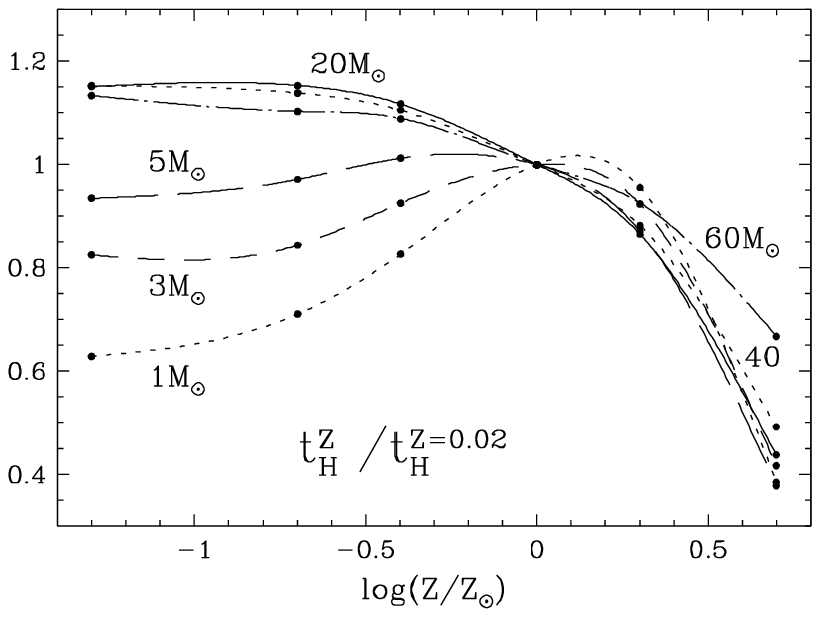}
\hfill
\includegraphics[width=2.6in,height=2.6in]{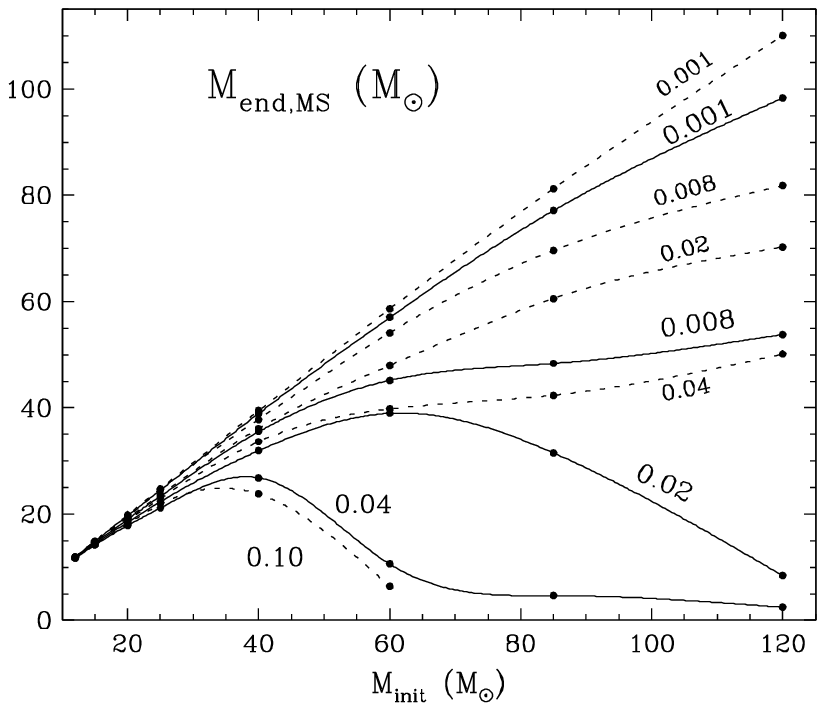}
\caption{{\it Left panels}: Main sequence lifetimes of models of initial masses as labeled on the curves, as a function of their metallicity. The lifetimes are normalized for each stellar mass to its value at $Z=0.02$. {\it Right panel}: Stellar masses at the end of the main sequence phase as a function
           of initial mass for different metallicities as labeled on the curves.
           Dotted lines correspond to models computed with moderate mass loss rates, while thick lines
           correspond to models computed with twice that 
           mass loss rates (see Mowlavi et al. 1998 for the references of the models used). Figures
           taken from Mowlavi et al. (1998b).}\label{fig3}
\end{figure}

The MS lifetimes are summarized in the left part of Fig.~\ref{fig3} for several stellar masses as a function of metallicity.
They can be understood in terms
of two factors: the initial H abundance, which determines the quantity of
available fuel, and the luminosity of the star, which fixes the rate at which
this fuel burns.

At $Z\leq 0.02$, $t_H$ is mainly determined by $L$, being shorter at higher
luminosities. The left part of Fig.~\ref{fig3} confirms, as expected from $L$ (see
left part of Fig.~\ref{fig1}), that $t_H$
increases with $Z$ in low- and intermediate-mass stars, and is about independent
of $Z$ in massive stars.

At $Z > 0.02$, on the other hand, the initial H abundance decreases sharply with
increasing $Z$ (the H depletion law is dictated by the adopted $\Delta Y/\Delta
Z$ law; for $\Delta Y/\Delta Z=2.4$ and $Y_0=0.24$, $X$ drops from 0.69 at $Z=0.02$
to 0.42 at $Z=0.1$). The MS
lifetimes are then mainly dictated by the amount of fuel available.
This, combined with the higher luminosities at $Z=0.1$, leads
to MS lifetimes which are about 60\% shorter at $Z=0.1$ than at
$Z=0.02$. This result is independent of the stellar mass for $M\leq 40$ M$_\odot$. Above
this mass, however, the action of mass loss (see below) extends the MS lifetime,
as can be seen from the 60 M$_\odot$ curve in the left part of Fig.~\ref{fig3}.

\subsection{The effect of $\dot{M}$}
\label{Sect:mass loss effect}

When mass loss is driven by radiation, Kudritzki et al. (1989, see also Vink et al. 2001)  showed that $\dot{M}$ in O stars is proportional to $Z^{0.5-0.8}$. As a
result, the effects of mass loss dominate in more metal-rich stars.
The stellar masses remaining at the end of the
MS phase for various mass loss rate prescriptions are shown in the left part of Fig.~\ref{fig3}.
The most striking result at metallicities higher than twice solar
is the rapid evaporation of massive stars with
$M \geq 50$ M$_\odot$, and the
consequent formation of WR stars during core H burning.

\subsection{The effects of rotation}
\label{Sect:rotation}

\begin{figure}
\includegraphics[width=2.6in,height=2.6in]{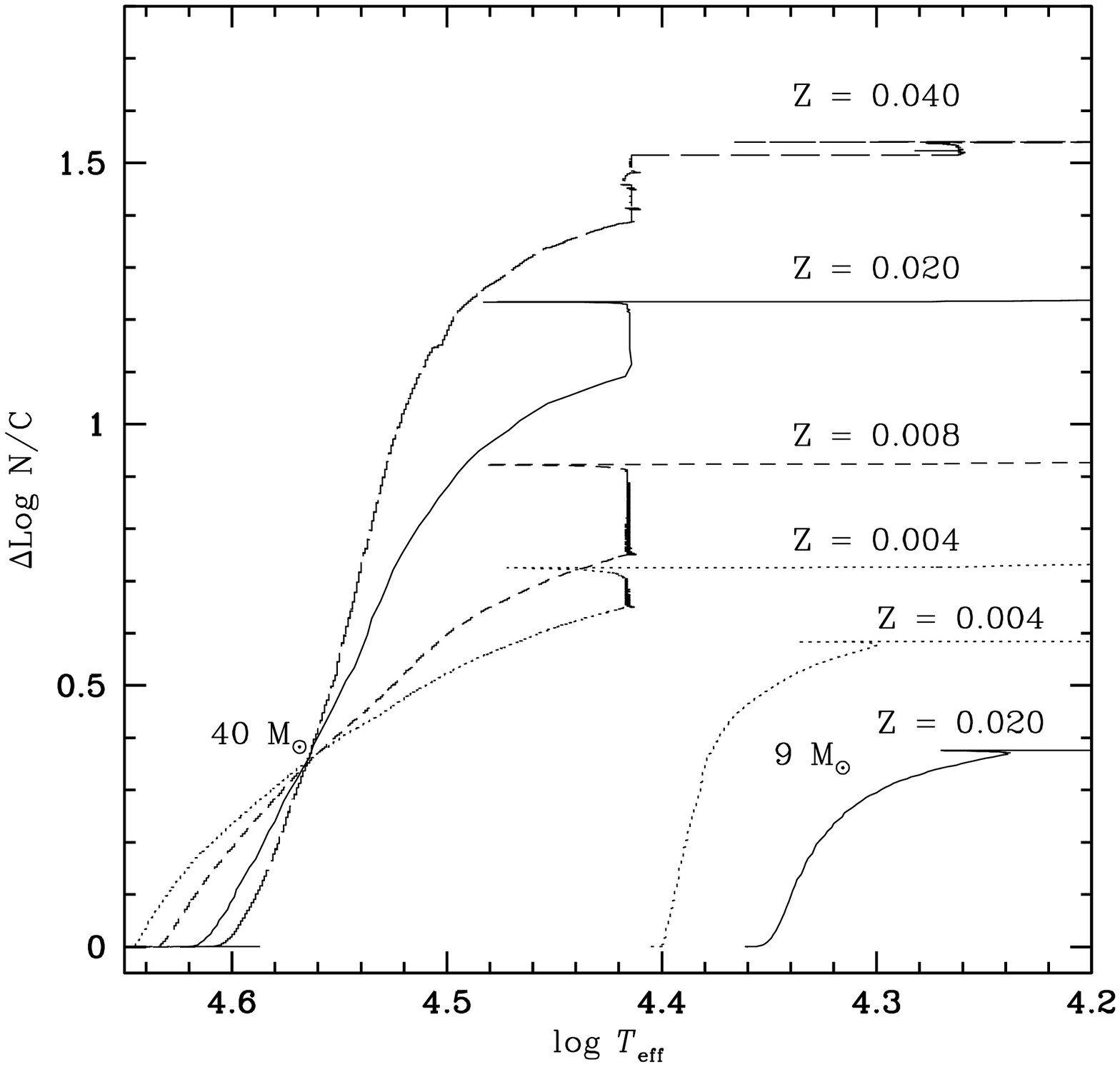}
\hfill
\includegraphics[width=2.6in,height=2.6in]{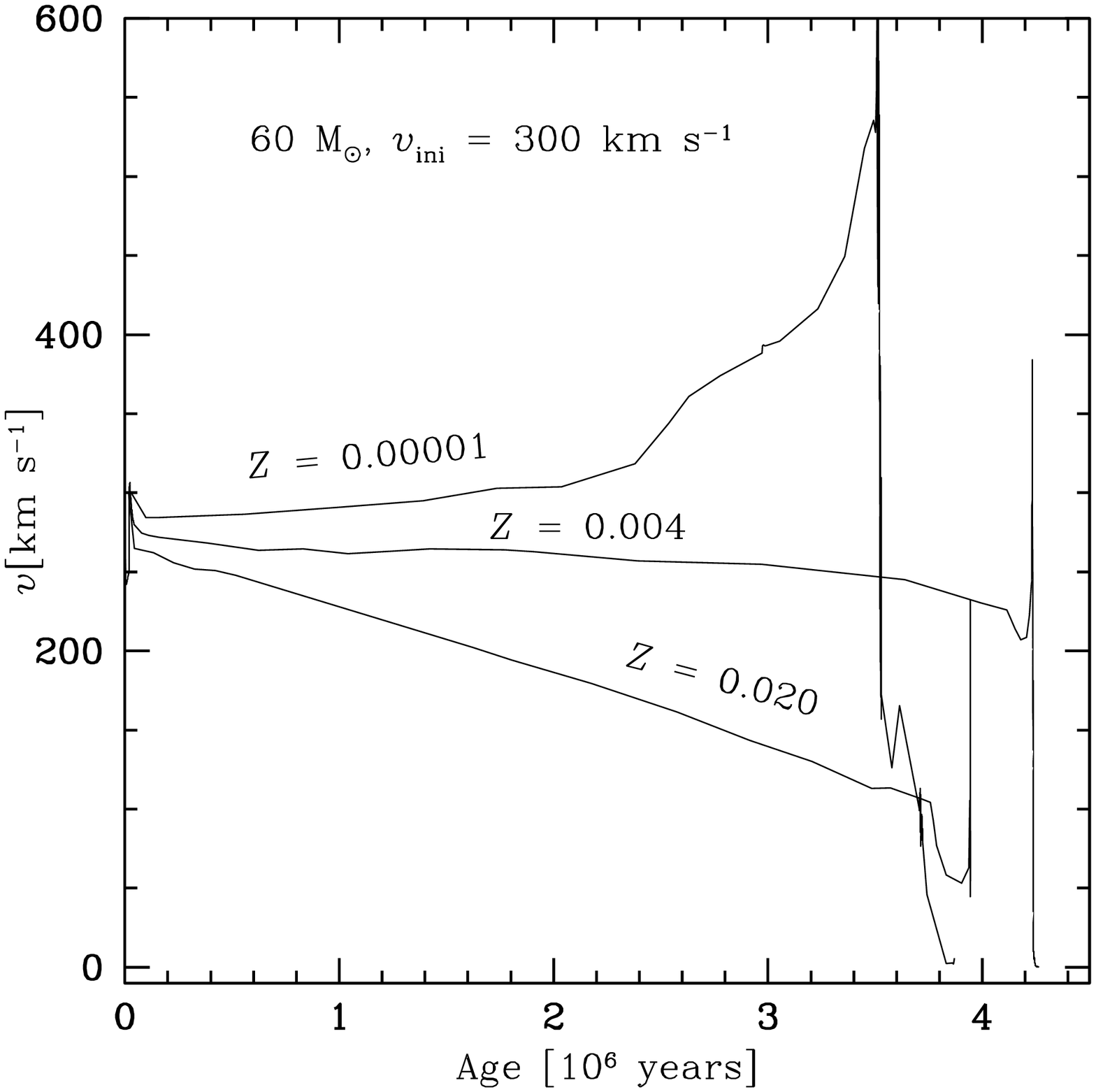}
\caption{{\it Left panels}: Evolution during the MS phase of the N/C ratios (in number) at the surface of 
  rotating stellar models as a function of the effective temperature. 
  The differences in N/C ratios are given with respect to the initial values.  {\it Right panel}: Evolution of the surface equatorial velocity as 
a function of time for 60 M$_\odot$ stars with $v_{\rm ini}$ = 300
km s$^{-1}$ at different initial metallicities.}\label{fig4}
\end{figure}

The effects of shellular rotation on the evolution of massive star models at high metallicity have been
discussed in Meynet \& Maeder (2005). As at lower metallicity, rotation induces mixing in the stellar interiors. However as was discussed in Maeder \& Meynet (2001), rotational mixing tends to be less efficient
in metal rich stars. This can be seen looking at the 9 M$_\odot$ stellar model in the left part of Fig.~\ref{fig4}.
This comes from the fact that
the gradients of $\Omega$ are much less steep in the higher metallicity
models, so they trigger less efficient shear mixing.
The gradients are shallower because 
less angular momentum is transported outwards by
meridional currents, whose velocity
scales as the inverse of the density in the outer layers
(see the Gratton-\"Opick term in the expression for the meridional velocity in Maeder \& Zahn~1998).
Looking at the 40 M$_\odot$ stellar model in the left part of Fig.~\ref{fig4}, one sees that the higher metallicity model presents the highest surface enrichments, in striking contrast with the behaviour of
the 9 M$_\odot$ model. This comes from the fact that
in the upper part of the HR diagram the evolution
of stars is more dominated by the mass loss rate than by rotation. And the changes occurring at the surface
of the 40 M$_\odot$ are thus not only due to rotation but also to mass loss which is more efficient at higher $Z$.

In the high mass range at high metallicity, the stars have less chance to reach the critical velocity during the Main Sequence phase as can be seen from the right part of Fig.~\ref{fig4}. This is due  to the fact that stellar winds remove the angular momentum at the surface, preventing thus the outer layers to reach the critical limit. 
Let us note that for smaller initial masses, a high metallicity may
favor the approach to the critical limit. Indeed, in that case, the stellar winds are too weak, even at high metallicity, to remove a lot of mass and therefore of angular momentum, while, the meridional currents,
which bring angular momentum from the inner regions of the star to the surface, are more rapid due to the
lower densities achieved in the outer layers of metal rich stars. Another point which should be kept in mind at this point is that, at very high metallicity, the stars may lose large amounts of mass without losing
too much angular momentum. This is due to the fact that in the high velocity regime, the stellar winds are 
polar (Maeder \& Meynet 2001). Let us stress however that this situation has little chance to be realised at
high metallicity. Indeed the timescales for mass loss are likely shorter than the timescale for the torque due to wind anisotropy to affect the surface velocity.

\section{Massive star population in high metallicity regions}

\begin{figure}
\includegraphics[width=2.6in,height=2.6in,angle=-90]{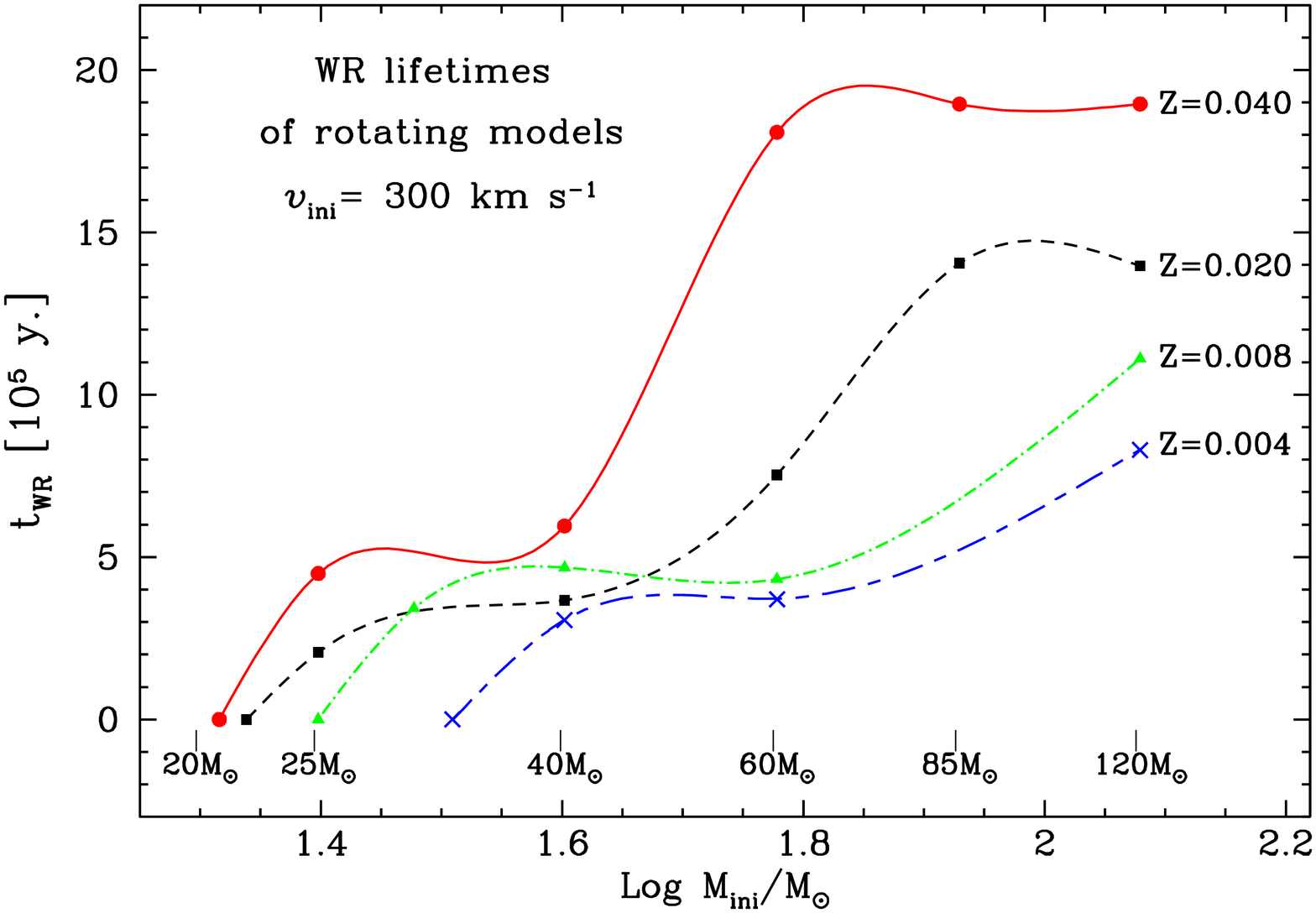}
\hfill
\includegraphics[width=2.6in,height=2.6in,angle=-90]{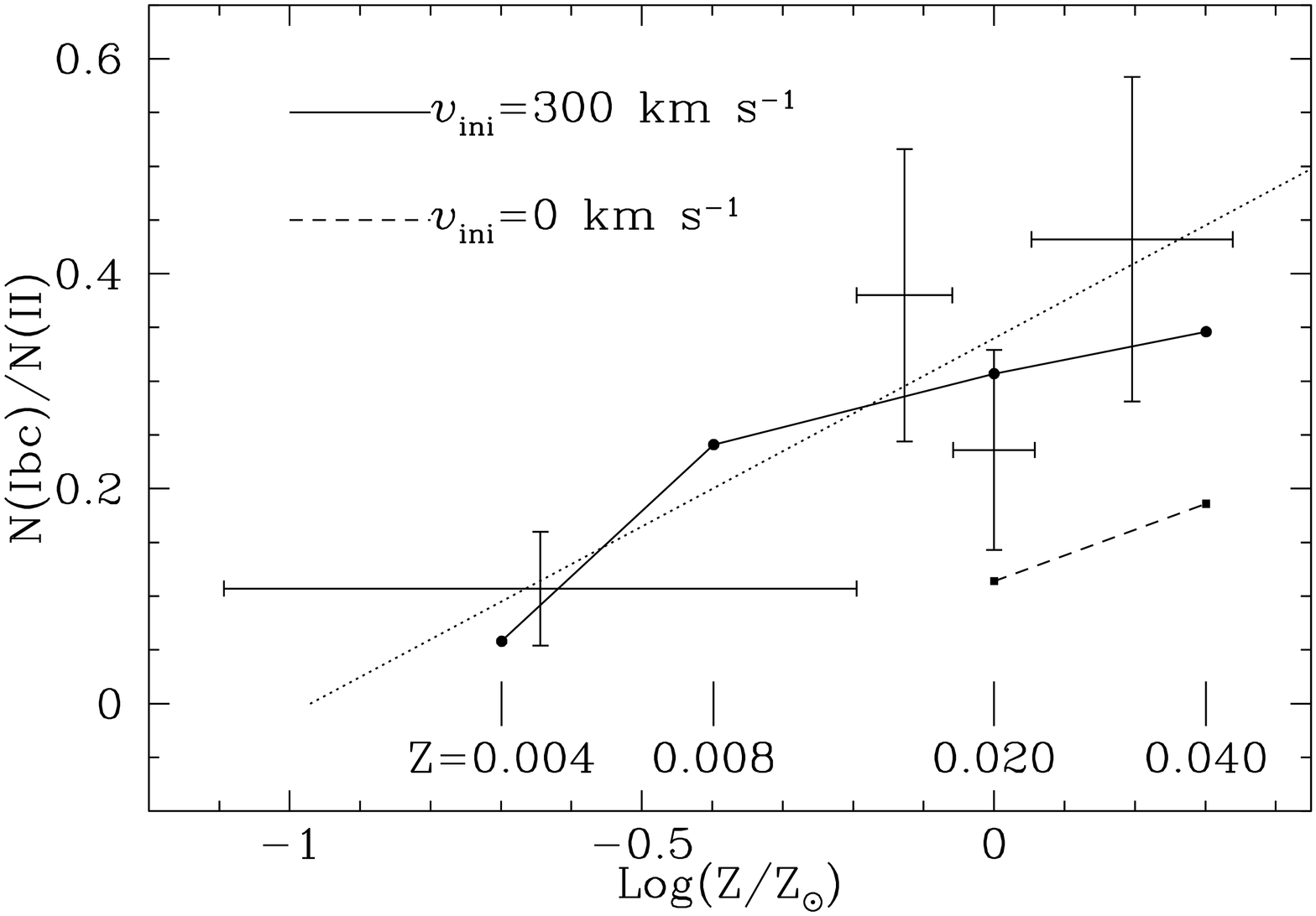}
\caption{{\it Left panels}: Lifetimes of Wolf--Rayet stars from various initial masses at four different metallicities.
All the models begin their evolution with $\upsilon_{\rm ini}$ = 300 km s$^{-1}$ on the ZAMS.  {\it Right panel}: Variation of the number ratios of type Ib/Ic supernovae to type II supernovae. The crosses
  with the error bars correspond to the values deduced from observations by Prantzos \& Boissier (2003). The
  dotted line is an analytical fit proposed by these authors. The continuous and dashed line show the predictions of  rotating and non--rotating stellar models. Note that
at high metallicity no type Ic supernovae are predicted by the models (see text). Figures taken from Meynet \& Maeder (2002) }\label{fig5}
\end{figure}

At high metallicity, as a result of the metallicity dependence of the mass loss rates, one expects larger
fractions of Wolf-Rayet (WR) stars (Maeder et al. 1980). This can be seen in the left part of Fig.~\ref{fig5} where the WR lifetimes of rotating models for four metallicities 
are plotted as a function of the initial mass.
The metallicity
dependence of the mass loss rates is responsible for two features: 1) for a given initial mass and velocity 
the WR lifetimes are greater at higher metallicities. Typically at Z=0.040
and for M $>$ 60 $M_\odot$ the WR lifetime is of the order of 2 Myr, while at the metallicity of the
SMC the WR lifetimes in this mass range are between 0.4--0.8 Myr; 2) the minimum mass 
for a single star to evolve into the WR phase is lower at higher metallicity. 

Comparisons with observed populations of WR stars are shown in Meynet \& Maeder (2005). 
When the variation with the metallicity of the number ratio of WR to O-type stars is considered,
good agreement is obtained provided models with rotation are used. Models without rotation predict much
too small fractions of WR to O-type stars, even at high metallicity. This illustrates the fact that
even at high metallicity, where the effects of the mass loss rates are dominant, one cannot neglect
the effects of rotation. 

Models well reproduce the variation with $Z$ of the WC/WN ratio at low metallicity, but underestimate this ratio at high $Z$ (see Fig.~11 in Meynet \& Maeder 2005). It might be that the mass loss rates during the post MS WNL phase are underestimated. Also, the metallicity dependence of the WR stellar winds
may help in resolving this disagreement (see Eldridge and Vink 2006).

Current knowledge associates supernovae of type Ib/Ic with the explosion of WR stars, the
H--rich envelope of which has been completely removed either by stellar winds and/or by mass transfer through Roche Lobe overflow in a close binary system. 
If we concentrate on the case of single star models, 
theory predicts that the fraction of supernovae progenitors without H-rich envelope
with respect to H-rich supernovae should be higher at higher metallicity. The reason is the same
as the one invoked to explain the increasing number ratio of WR to O--type stars with metallicity, namely the
growth of the mass loss rates with $Z$. Until recently very little observational evidence has been found 
confirming this predicted behaviour. The situation began to change with the work by Prantzos \& Boissier~(2003). These authors
have derived from published data the
observed number ratios of type Ib/Ic supernovae to
type II supernovae for different metallicities. The regions considered  are regions of constant star formation rate. 
Their results are plotted in the right part of Fig.~\ref{fig5}.

Looking at this figure, it clearly appears that rotating models
give a much better fit to the observed data than non--rotating models. This comparison can be viewed as a check
of the lower initial mass limit M$_{\rm WNE}$ of the stars evolving into a WR phase without hydrogen. Let us note also at this point that at high metallicity, no type Ic supernovae are expected to occur (supernovae
showing no trace of H and He). This is a consequence of the high mass loss which allows the star to enter
at an early stage into the WC phase when still a lot of helium has to be transformed into carbon and oxygen.
The star thus keeps a high abundance of helium at its surface until the presupernova stage and explodes as a
type Ib supernova. At low metallicities, since the mass loss rates are weaker, the star may enter (if it does)
at a later stage of the core He-burning phase, when already most of the helium has been transformed into
carbon and oxygen (see Smith \& Maeder 1991). In that case the star may explode as a type Ic supernova. This 
might explain why long soft Gamma Ray Bursts associated to type Ic supernovae only occur in metal poor regions
(Hirschi et al. 2006).

\begin{figure}
\includegraphics[width=3.4in,height=2.0in]{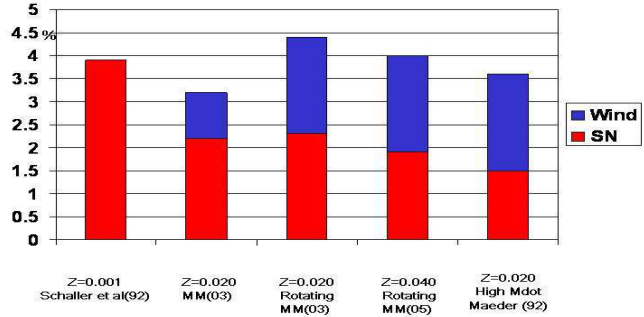}
\caption{Mass of new elements ejected by stars more massive than 8 M$_\odot$
per mass in stars given by different stellar models. A Salpeter IMF has been used. The labels MM03 and MM05
are for respectively Meynet \& Maeder 2003 and Meynet \& Maeder 2005.}\label{fig6}
\end{figure}

\begin{figure}
\includegraphics[width=4.4in,height=3.0in]{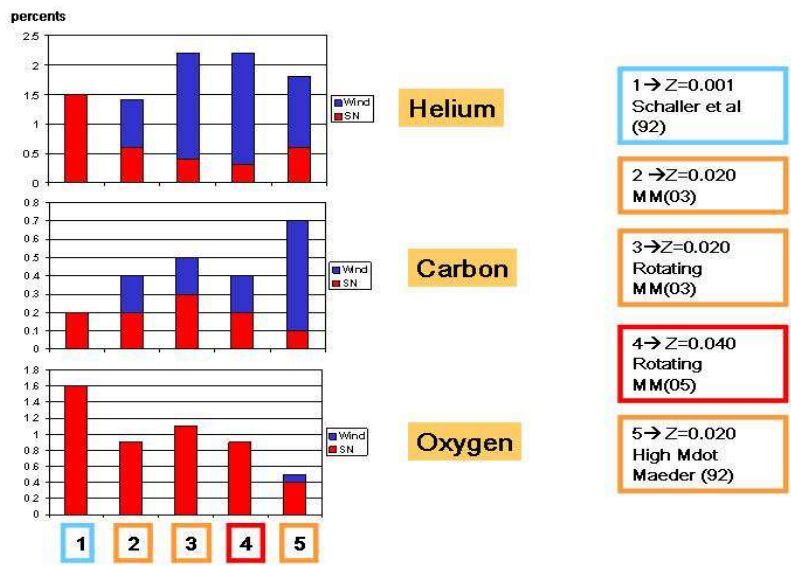}
\caption{Mass of new helium, carbon and oxygen ejected by stars more massive than 8 M$_\odot$
per mass initially in stars given by different stellar models. A Salpeter IMF has been used. The labels MM03 and MM05
are for respectively Meynet \& Maeder 2003 and Meynet \& Maeder 2005. }\label{fig7}
\end{figure}

\section{Massive star nucleosynthesis in high metallicity regions}

Let us begin this section by recalling a few orders of magnitudes. When single stars form with a Salpeter's IMF,
about 14\% of the mass locked into stars consists of massive stars, {\it i.e.} with masses greater than 8 M$_\odot$, 25\%  are locked into stars of masses between 1 and 8 M$_\odot$ and 61\% in stars
with masses between 0.1 and 1 M$_\odot$. When all the stars with masses above 1 M$_\odot$ have died,
about 13\% of the mass initially locked into stars is ejected by massive stars (1\% remains locked into black holes or neutron stars). The intermediate mass stars (from 1 to 8 M$_\odot$) eject about 18.5\% of the mass initially locked into stars (6.5\% remains locked in white dwarfs).

Figure 6 shows for various models at different metallicities the fraction of the mass in stars which is
eventually ejected under the form of new elements. The part ejected by stellar winds is distinguished from that
ejected at the time of the supernova explosion. One sees that on the whole, the fraction of the mass initially locked into stars and transformed into new elements by the massive stars does not depend too much on the model or on the metallicity. All the results are comprised between 3.5 and 4.5\%. 

One notes however that more variations appear when one looks at the proportions of these new elements ejected
by the winds and the supernova explosion. Indeed at higher metallicity a greater part of the new elements synthesized by stars are ejected by stellar winds
as can be see in Fig~\ref{fig6}. Typically the non-rotating stellar models of Schaller et al. (1992)
predict that a stellar population at $Z=0.001$ ejects in the form of new elements
during the supernova explosion a little less than 4\% of the total mass used to form the stars. Models at solar metallicity and higher eject
about half of their new elements at the time of the supernova explosion and the rest through stellar winds.
This may have a great influence on the final yields as has been shown by Maeder (1992). Indeed, mass loss removes matter at earlier evolutionary stages. The ejected matter has therefore been partially processed by nuclear burning and has a chemical composition different from the one it would have if the matter would have remained locked in the star.

This effect is responsible for many specific enrichments at high metallicity: for instance, because
of this effect, massive star
models are expected to be stronger sources of $^{4}$He, $^{12}$C, $^{22}$Ne, $^{26}$Al and to be less 
important sources of $^{16}$O at high metallicity than at low metallicity. The cases of  helium, carbon and oxygen is shown in Fig~\ref{fig7}. We can see the importance of mass loss for the carbon yields. Indeed the 
larger yields are obtained for the models computed with enhanced mass loss rates. It is interesting here to
note that high mass loss is not a sufficient condition for obtaining high carbon yields. High carbon yields are obtained only when the star enters into the WC phase at an early stage of the core He-burning phase.
This can be seen 
comparing the 60 M$_\odot$ stellar model at $Z=0.04$ computed by Meynet \& Maeder (2005) with the 60 M$_\odot$ stellar model at $Z=0.02$ with enhanced mass loss rate computed by Maeder (1992).
These two models end their lifetimes with similar final masses (the model of Meynet \& Maeder 2005 with 6.7 M$_\odot$ and the model of Maeder 1992 with 7.8 M$_\odot$), thus the mass of new carbon ejected at the time
of the supernova explosion is quite similar for both models and is around 0.45 M$_\odot$. When one
compares the mass of new carbon ejected by the winds, we obtain 7.3 and 0.16  M$_\odot$ for respectively the
model of Maeder (1992) and the one of Meynet \& Maeder (2005). This difference arises from the fact that the model
with enhanced mass loss enters the WC phase at a much earlier time of the core He-burning phase, typically when
the mass fraction of helium at the centre, $Y_c$ is 0.43 and the actual mass of the star is 25.8 M$_\odot$,
while the model of Meynet \& Maeder (2005) enters the WC stage when $Y_c$ is 0.24 and the actual mass 8.6 M$_\odot$. 

The example above shows that the entry at an early stage into the WC phase is more favored
by strong mass loss than by rotation. This comes from the fact that rotation favors an early entry into the
WN phase, while the star has still an important H-rich envelope. It takes time for the whole H-rich envelope to be removed and when it is done, the star is already well advanced into the core He-burning phase.
Of course this conclusion is quite dependent on the magnitudes of the mass loss rates. For instance higher
mass loss rates during (a part of) the WNL phase would favor an early entrance into the WC phase. This would give a better agreement with the number ratio of WC to WN observed and would lead to higher carbon yields.

Let us conclude this paper by estimating an empirical yield in carbon from the Wolf-Rayet stars in the solar neighborhood. From the catalogue by van der Hucht (2001) one obtains that the number of WC stars at a distance less than 3 kpc around the Sun is 44. The mass loss rate during the WC phase is estimated to
be 10$^{-4.8}$ M$_\odot$ per year, the mass fraction of carbon observed in the WC stellar wind is around 0.35 (see Table 2 in the review by Crowther 2006). If we
consider a star formation rate of 2-4 M$_\odot$ per square parsec and per Gyr, one obtains that the mass of (new) carbon ejected by WC wind per mass used to form stars is between 0.25 and 0.5\%. These empirical yields are well in lines with the range of values given by the models 
which are between 0.2-0.6\% (see Fig.~\ref{fig7}, the greatest value corresponding to the case of Maeder 1992). Let us note that incorporating the yields of Maeder (1992) into chemical evolution models has an important impact (see e.g. Prantzos et al. 1994). If the upper value of the empirical yields is the correct one then this indicates that
WC stars are very important sources of carbon in metal rich regions.

\end{document}